\documentclass{jfm}
\usepackage{Paper}

\shorttitle{}
\shortauthor{N. Nair and A. Goza}

\title[Hybrid flow control using reinforcement learning]{\Large Bio-inspired variable-stiffness flaps for hybrid flow control, tuned via reinforcement learning}

\author{Nirmal J. Nair\aff{1}
  \corresp{\email{njn2@illinois.edu}}
  \and Andres Goza\aff{1}}

\affiliation{\aff{1}Department of Aerospace Engineering, University of Illinois Urbana-Champaign, IL 61801, USA}

\newcommand{\RR}{\ensuremath{\mathbb{R}}} 

\newcommand{\state}[1]{\ensuremath{\bm{s}_{#1}}}
\newcommand{\action}[1]{\ensuremath{a_{#1}}}
\newcommand{\reward}[1]{\ensuremath{r_{#1}}}
\newcommand{\timecontrol}{\ensuremath{m}}

\newcommand{\rltimesteps}{\ensuremath{M_{\tau}}}
\newcommand{\nstates}{\ensuremath{N_{s}}}
\newcommand{\nfsi}{\ensuremath{N_{t}}}
\newcommand{\trajectory}{\ensuremath{\tau}}
\newcommand{\ntrajectories}{\ensuremath{N_{\tau}}}
\newcommand{\discountfactor}{\ensuremath{\gamma}}
\newcommand{\std}{\ensuremath{\sigma}}

\newcommand{\vorticity}{\ensuremath{\bm{\omega}}}

\newcommand{\nepochs}{\ensuremath{n_{e}}}

\newcommand{\episodemax}{\ensuremath{M_i}}
\newcommand{\multiple}{\ensuremath{k}}

\newcommand{\penalone}{\ensuremath{p_1}}
\newcommand{\penaltwo}{\ensuremath{p_2}}
\newcommand{\undeployedcurrent}{\ensuremath{u}}
\newcommand{\undeployedmax}{\ensuremath{u_{max}}}

\newcommand{\initialstatespace}{\ensuremath{\mathcal{S}^0}}

\newcommand{\expected}{\ensuremath{\mathbb{E}}}

\newcommand{\objective}{\ensuremath{J}}

\newcommand{\clip}{\ensuremath{\epsilon}}

\newcommand{\rlweights}{\ensuremath{\bm{\theta}}}

\newcommand{\policy}{\ensuremath{\pi_{\rlweights}}}

\newcommand{\counter}{\ensuremath{count}}

\newcommand{\chord}{\ensuremath{c}}

\newcommand{\reynolds}{\ensuremath{Re}} 
\newcommand{\velocityscale}{\ensuremath{U_{\infty}}} 
\newcommand{\defl}{\ensuremath{\beta}}
\newcommand{\deflection}{\ensuremath{\beta}}
\newcommand{\stiff}{\ensuremath{k_{\defl}}}

\newcommand{\mass}{\ensuremath{m_{\defl}}}
\newcommand{\sixty}{\ensuremath{60\%}}
\newcommand{\timet}{\ensuremath{t}}
\newcommand{\periodt}{\ensuremath{t/T}}

\newcommand{\lev}{\ensuremath{\Gamma_{LEV}}}

\newcommand{\tev}{\ensuremath{\Gamma_{TEV}}}

\newcommand{\liftmean}{\ensuremath{\overline{C}_l}}

\newcommand{\Cp}{\ensuremath{C_p}}
\begin{document}

\maketitle

\begin{abstract}
A bio-inspired, passively deployable flap attached to an airfoil by a torsional spring of fixed stiffness can provide significant lift improvements at post-stall angles of attack. In this work, we describe a hybrid active-passive variant to this purely passive flow control paradigm, where the stiffness of the hinge is actively varied in time to yield passive fluid-structure interaction (FSI) of greater aerodynamic benefit than the fixed-stiffness case. This hybrid active-passive flow control strategy could potentially be implemented using variable stiffness actuators with less expense compared with actively prescribing the flap motion. The hinge stiffness is varied via a reinforcement learning (RL)-trained closed-loop feedback controller. A physics-based penalty and a long-short-term training strategy for enabling fast training of the hybrid controller are introduced. The hybrid controller is shown to provide lift improvements as high as 136\% and 85\% with respect to the flap-less airfoil and the best fixed-stiffness case, respectively. These lift improvements are achieved due to large-amplitude flap oscillations as the stiffness varies over four orders of magnitude, whose interplay with the flow is analyzed in detail.

\end{abstract}



\section{Introduction}

For aerodynamic flows, a passive flow control device (flap) inspired from self-actuating covert feathers of birds has been shown to improve lift at post-stall angles of attack \citep{bramesfeld2002experimental,duan2021covert}. In particular, when the flap is mounted via a torsional spring, further aerodynamic benefits can be obtained compared with a free (zero hinge stiffness) or static configuration \citep{rosti2018passive}. These added benefits arise from rich fluid-structure interaction (FSI) between the flap and vortex dynamics \citep{nair2022fluid}. This outcome teases a question: can additional lift enhancement be achieved if the flap motion was controlled to yield more favorable flapping amplitudes and phase relative to key flow processes?

To address this question, we propose a hybrid active-passive flow control method to adaptively tune the flap stiffness. That is, the flap dynamics are \emph{passively} induced by the FSI, according to the \emph{actively} modulated hinge stiffness. This hybrid approach could incur less expense as compared to a fully active control method where the flap deflection is controlled using a rotary actuator. Our focus is on the design of a control algorithm that can actuate the hinge stiffness to provide aerodynamic benefits without accounting for how these stiffness changes are implemented, and on explaining the physical mechanisms that drive these benefits. We note, however, that  
there are various ways of achieving stiffness modulation in practice via continuous variable stiffness actuators (VSA) \citep{wolf2015variable}, used extensively in robotics \citep{ham2009compliant}, wing morphing \citep{sun2016morphing}, \emph{etc}. 
Discrete VSA restricts the stiffness to vary discretely across fixed stiffness levels but it weighs less and requires lower power \citep{diller2016lightweight}. 
 


Historically, linear approximations of fundamentally nonlinear systems are used to design optimal controllers \citep{kim2007linear}. 
While these linear techniques have been effective in stabilizing separated flows at low $Re\sim \mathcal{O}(10^2)$ where the base state has a large basin of attraction, its effectiveness is compromised at larger $Re$ \citep{ahuja2010feedback}. These challenges are exacerbated by the nonlinear FSI coupling between the flap and vortex shedding of interest here.
Model predictive control (MPC) uses nonlinear models to make real-time predictions of the future states to guide the control actuations. The need for fast real-time predictions necessitates the use of reduced-order models where the control optimization problem is solved using a reduced system of equations \citep{peitz2020data}. Machine learning in fluid mechanics has provided further avenues of deriving more robust reduced nonlinear models to be used with MPC \citep{bieker2020deep,baumeister2018deep,mohan2018deep}. However, these reduced-order modeling efforts remain an area of open investigation, and would be challenging for the strongly coupled flow-airfoil-flap FSI system. We therefore utilize a model-free, reinforcement learning (RL) framework to develop our controller. RL has recently gained attention in fluid mechanics \citep{garnier2021review}, and is used to learn an effective control strategy by trial-and-error via stochastic agent-environment interactions \citep{sutton2018reinforcement}. 
Unlike MPC, the control optimization problem is completely solved offline, thereby not requiring real-time predictions. RL has been successfully applied to attain drag reduction \citep{rabault2019artificial,paris2021robust,fan2020reinforcement,li2022reinforcement}, shape optimization \citep{viquerat2021direct} and understanding swimming patterns \citep{verma2018efficient,zhu2021numerical}. 

In this work, we develop a closed-loop feedback controller using deep RL for our proposed hybrid control approach consisting of a tunable-stiffness covert-inspired flap. We train and test this controller using high-fidelity fully coupled simulations of the airfoil-flap-flow dynamics, and demonstrate the effectiveness of the variable-stiffness control paradigm compared with the highest performing passive (single-stiffness) case. We explain the lift-enhancement mechanisms by relating the large-amplitude flap dynamics to those of the vortex formation and shedding processes around the airfoil. 


\section{Methodology}
\subsection{Hybrid-active-passive control}
\label{hybridcontrol}



\begin{figure}
\centering
\includegraphics[scale=1]{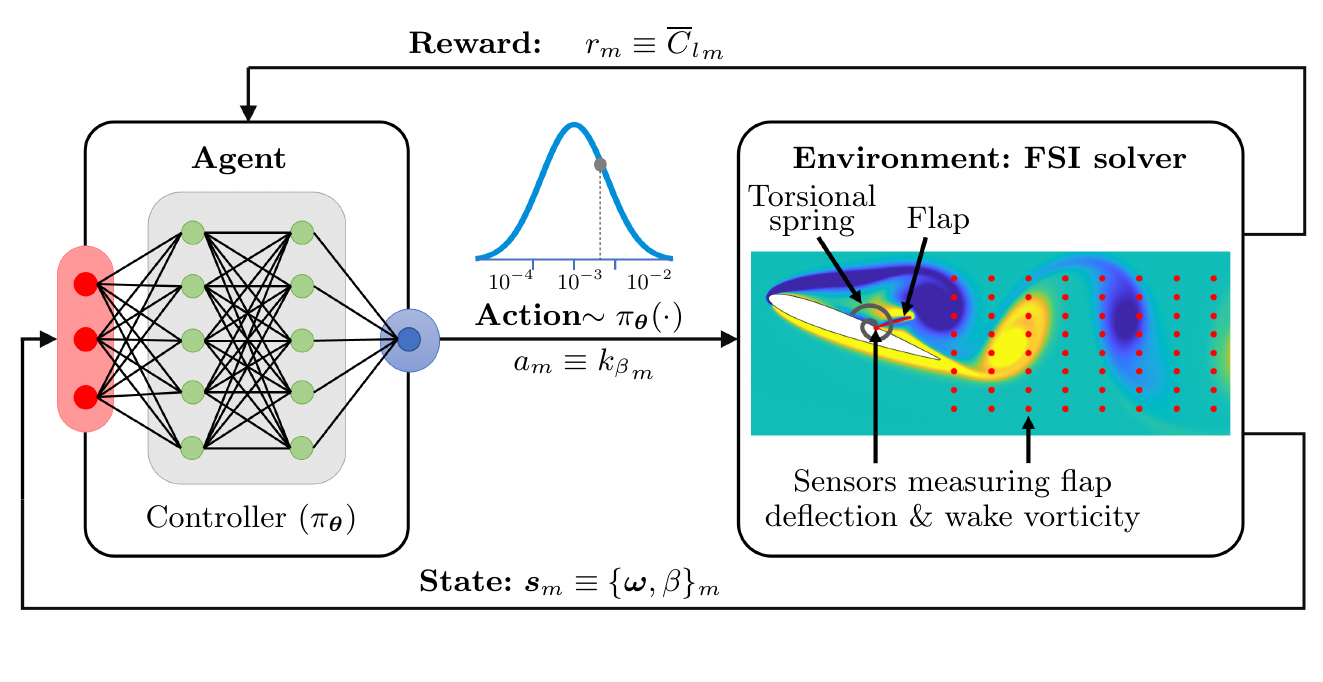}
\vspace{-0.9cm}
\caption{Schematic of the problem setup and RL framework.}
\label{nnprob}
\end{figure}

The problem setup is shown in Fig.~\ref{nnprob}, which consists of a NACA0012 airfoil of chord $c$ at an angle of attack of $20^\circ$ and $\reynolds = 1 {,}000$, where significant flow separation and vortex shedding occur. A flap of length $0.2c$ is hinged on the upper surface of the airfoil via a torsional spring with stiffness, $\stiff$, where $\deflection$ denotes the deflection of the flap from the airfoil surface. 
In the passive control approach \citep{nair2022fluid}, $\stiff$ was fixed and maximum lift was attained at $\stiff=0.015$. In our hybrid active-passive control the stiffness is a function of time, $\stiff(t)$, determined by a RL-trained closed-loop feedback controller described in Sec.~\ref{rlgeneral}. While the stiffness variation is allowed to take any functional form, it is restricted to vary in $\stiff(t) \in [10^{-4}, 10^{-1}]$, similar to the range of stiffness values considered in the passive control study. The mass and location of the flap are fixed at $\mass=0.01875$ and $\sixty$ of the chord length from the leading edge, chosen here since they induced the maximal lift benefits in the passive (single-stiffness) configuration (\emph{c.f.}, Fig.~\ref{fvv1}--\ref{fvv4} for vorticity contours at four time instants in one periodic lift cycle for this highest-lift single-stiffness case). 


\subsection{Reinforcement learning (RL)}
\label{rlgeneral}





A schematic of the RL framework is shown in Fig. \ref{nnprob} where an agent (with a controller) interacts with an environment. At each time step, $\timecontrol$, the agent observes the current state of the environment, $\state{\timecontrol} \in \RR^{\nstates}$, where $\nstates$ is the number of states, implements an action, $\action{\timecontrol} \in \RR$, and receives a reward, $\reward{\timecontrol} \in \RR$. The environment then advances to a new state, $\state{\timecontrol+1}$. This process is continued for $\rltimesteps$ time steps and the resulting sequence forms a trajectory, $\trajectory= \{\state{0},\action{0},\reward{0},\ldots,\state{\rltimesteps},\action{\rltimesteps},\reward{\rltimesteps}\}$. 
The actions chosen by the agent to generate this trajectory are determined by the stochastic policy of the controller, $\policy(\action{\timecontrol}|\state{\timecontrol})$, parametrized by weights, $\rlweights$. This policy outputs a probability distribution of actions, from which $\action{\timecontrol}$ is sampled, $\action{\timecontrol} \sim \policy(\action{\timecontrol}|\state{\timecontrol})$, as shown in Fig.~\ref{nnprob}. In policy-based deep RL methods, a neural network is used as a nonlinear function approximator for the policy as shown in Fig.~\ref{nnprob}. Accordingly, $\rlweights$ corresponds to the weights of the neural network.
The goal in RL is to learn an optimal control policy that maximizes an objective function $\objective(\cdot)$, defined as the expected sum of rewards as
\begin{equation}
\objective(\rlweights) = \expected_{\trajectory \sim \policy} \left[ \sum_{m=0}^{\rltimesteps} \reward{\timecontrol} \right].
\label{rewardssum2}
\end{equation}
Here, the expectation is performed over $\ntrajectories$ different trajectories sampled using the policy, $\trajectory \sim \policy$. The maximization problem is then solved by gradient ascent where 
an approximate gradient of the objective function, $\nabla_{\rlweights} \objective(\rlweights)$, is obtained from the policy gradient theorem \citep{nota2019policy}. 
In this work, we use the PPO algorithm \citep{schulman2017proximal} for computing the gradient, which is a policy-based RL method suitable for continuous control problems as opposed to $Q$-learning methods for discrete problems \citep{sutton2018reinforcement}. PPO has been used successfully to develop RL controllers for fluid flows  \citep{rabault2019artificial}, and is chosen here among other policy-based methods due to its relative simplicity in implementation, better sample efficiency, and ease of hyperparameter tuning. 






In our hybrid control problem, the environment is the strongly-coupled FSI solver of \cite{nair2022strongly}, where the incompressible Navier-Stokes equation for the fluid coupled with Newton's equation of motion for the flap are solved numerically. 
The state provided as an input to the controller are sensor measurements consisting of flow vorticity in the wake, $\vorticity_\timecontrol$, and flap deflection, $\deflection_\timecontrol$. 
The action is the time-varying stiffness, $\action{\timecontrol} = {\stiff}_{\timecontrol} \in \RR^+$. 
Similar to \cite{rabault2019artificial}, when advancing a single \emph{control} time step from $\timecontrol$ to $\timecontrol+1$, the flow-airfoil-flap system is simulated for $\nfsi$ \emph{numerical} time steps of the FSI solver. In this duration, the chosen value of the stiffness is kept constant. The reason for introducing these two time scales---control and numerical---is to allow the FSI system to meaningfully respond to the applied stiffness and achieve faster learning. 



The reward for the lift maximization problem of our hybrid control approach is
\begin{equation}
\reward{\timecontrol} = \frac{1}{2}{\liftmean}^2_m + \penalone \left( \frac{1-\penaltwo^{\undeployedcurrent/\undeployedmax}}{\penaltwo}\right).
\label{rewpenal}
\end{equation}
The first term is the mean lift coefficient of the airfoil, ${\liftmean}_m$, evaluated over the $\nfsi$ numerical time steps. 
The second term, where $\penalone>0$, $\penaltwo \gg 1$ are constants whose values are given in Sec.~\ref{rlresults}, is a physics-based penalty term that provides an exponentially growing negative contribution to the reward if the flap remains undeployed for several consecutive control time steps (intuitively, one wishes to avoid periods of prolonged zero deployment angle). Accordingly, $\undeployedcurrent$ denotes the current count of consecutive control time steps that the flap has remained undeployed and $\undeployedmax$ is the maximum number of consecutive time steps that the flap may remain undeployed. The flap is deemed undeployed if $\deflection_{\timecontrol} < \deflection_{min}$.



\begin{algorithm}[b!]
\footnotesize
\caption{PPO-RL applied to hybrid control}\label{pporl}
  \begin{algorithmic}[1]
    \REQUIRE Set of initial conditions, \initialstatespace
    \ENSURE Optimization parameters, $\rlweights$
    \STATE Initialize state, $\state{0} \sim \initialstatespace$
    \STATE Initiate a vector of counters for each trajectory, $\undeployedcurrent[1:\ntrajectories]\leftarrow 0$, $\counter[1:\ntrajectories]\leftarrow 0$
    \FOR{$\text{iterations} \leftarrow 1,2,\ldots$}
    \FOR[multi-window optimization]{$\text{window} \leftarrow 1,2,\ldots, \multiple$ \label{algokwindow}}
    \FOR[perform trajectory sampling]{$\timecontrol \leftarrow 1,2,\ldots, \rltimesteps(=\episodemax/\multiple)$ \label{algotraj}}
    \FOR[parallel trajectory sampling]{$\text{trajectory}:j \leftarrow 1,2,\ldots, \ntrajectories$ \label{algoparallel}}
    \STATE $\action{\timecontrol}[j]\sim \policy(\action{\timecontrol}[j]| \state{\timecontrol}[j])$
    \STATE $\reward{\timecontrol}[j], \state{\timecontrol+1}[j] = FSI(\state{\timecontrol}[j],\action{\timecontrol}[j])$ \label{algofsi}
    \STATE $\counter[\text{j}] \leftarrow \counter[\text{j}] + 1$
    \IF{$\deflection_{\timecontrol+1}[j] < \deflection_{min}$}
    \STATE $\undeployedcurrent[\text{j}] \gets \undeployedcurrent[\text{j}] + 1$
    \ENDIF
    \IF[episode termination]{$\undeployedcurrent[\text{j}]=\undeployedmax$ $||$ $\counter[\text{j}]=\episodemax$ \label{algoepterm}}
    \STATE Reset to initial condition, $\state{\timecontrol+1}[j] \sim \initialstatespace$ \label{algoreset}
    \STATE $\undeployedcurrent[\text{j}]\leftarrow 0$, $\counter[\text{j}]\leftarrow 0$
    \ENDIF
    \ENDFOR
    \ENDFOR
    \STATE Optimize $\rlweights$ using gradient ascent.
    \label{algooptimize}
    \ENDFOR
    \ENDFOR
  \end{algorithmic}
\end{algorithm}

The RL algorithm is proceeded iteratively as shown in Algorithm \ref{pporl}. Each iteration consists of sampling trajectories spanning a total of $\episodemax$ control time steps (lines \ref{algokwindow} and \ref{algotraj}) and using the collected data to optimize $\rlweights$ (line~\ref{algooptimize}). We also define an episode as either the full set of $\episodemax$ time steps or, in the case where the parameters yield an undeployed flap, the time steps until $\undeployedcurrent=\undeployedmax$. Note that an episode and iteration coincide only if an episode is not terminated early. The state is only reset after an episode terminates (line~\ref{algoreset}), which could occur within an iteration if the episode terminates early (line~\ref{algoepterm}). 
We also use a modified strategy to update $\rlweights$ in a given iteration. 
In policy-based methods, the weights update step is performed after trajectory sampling. Generally, in one iteration, trajectories are collected only once. This implies that (a) typically the weights are updated once in one training iteration and (b) the length of the trajectory is equal to the iteration length, $\rltimesteps = \episodemax$.  
However, in our work, we perform $\multiple>1$ number of weight updates in a single iteration by sampling $\multiple$ trajectories each of length $\rltimesteps = \episodemax/ \multiple$. 
This procedure is found to exhibit faster learning since the frequent weight updates sequentially cater to optimizing \emph{shorter} temporal windows of the \emph{long}-time horizon. We therefore refer to this procedure as the long-short-term training strategy and demonstrate its effectiveness in Sec.~\ref{rlresults}. As shown in Algorithm~\ref{pporl}, each iteration is divided into $\multiple$ optimization windows (line~\ref{algokwindow}) and $\rlweights$ is updated at the end of each window (line~\ref{algooptimize}). Finally, for computing more accurate estimates of the expected values used in Eq.~\ref{rewardssum2} and for evaluating gradients, the $\timecontrol$th time advancement (line~\ref{algotraj}) is performed $\ntrajectories$ times independently (line~\ref{algoparallel}) \citep{schulman2017proximal}. For accelerated training, this set of $\ntrajectories$ trajectories are sampled in parallel \citep{rabault2019accelerating,pawar2021distributed}.
\section{Results}
\label{rlresults}



\subsection{RL and FSI parameters}

The parameters of the FSI environment are the same as in \cite{nair2022strongly}, which contains the numerical convergence details. The spatial grid and time step sizes are $\Delta x/\chord=0.00349$ and $\Delta t/(\chord/\velocityscale) = 0.0004375$, respectively. For the multi-domain approach for far-field boundary conditions, five grids of increasing coarseness are used where the finest and coarsest grids are $[-0.5,2.5]\chord \times [-1.5, 1.5]\chord$ and $[-23,25]\chord \times [-24, 24]\chord$, respectively. The airfoil leading edge is located at the origin. For the sub-domain approach, a rectangular sub-domain that bounds the physical limits of flap displacements, $[0.23,0.7] c\times [-0.24,0.1] c$, is utilized. The FSI solver is parallelized and simulated across six processors.

For the states, $\nstates = 65$ sensor measurements are used which measure vorticity at $64$ locations distributed evenly across $[1,2.4] c\times [-0.6,0.1] c$ and flap deflection as denoted by the red markers in Fig.~\ref{nnprob}. 
To ensure unbiased stiffness sampling across $[10^{-4}, 10^{-1}]$, a transformation between stiffness and action is introduced: ${\stiff}_{\timecontrol} = 10^{\action{\timecontrol}}$. Accordingly, $\action{\timecontrol}$ is sampled from a normal distribution, $\mathcal{N}(\action{\timecontrol},\std)$ in the range $[-4, -1]$, so that ${\stiff}_{\timecontrol}$ is sampled from a log-normal distribution. 
The neural network consists of fully-connected layers with two hidden layers. The size of each hidden layer is $64$ and the hyperbolic tangent (tanh) function is used for nonlinear activations. 
The parameters in the reward function \eqref{rewpenal} are $\penalone=0.845$, $\penaltwo=10,000$, $\undeployedmax=20$ and $\deflection_{min}=5^\circ$.

The initial state for initializing every episode corresponds to the limit cycle oscillation solution obtained at the end of a simulation with a constant $\stiff=0.015$ spanning $40$ convective time units ($\timet=0$ in this work denotes the instance at the end of this simulation).  In advancing one control time step, $\nfsi=195$ numerical time steps of the FSI solver are performed, or approximately $0.085$ convective time units. That is, the control actuation is provided every $5\%$ of the vortex-shedding cycle. 
The various PPO-related parameters are the discount factor of $\discountfactor=0.9$, learning rate of $\alpha=0.0003$, $\nepochs=10$ epochs and clipping fraction of $\clip=0.2$. Refer to \cite{schulman2017proximal} for the details of these parameters. 
$\ntrajectories=3$ trajectories are sampled in parallel.  
The PPO-RL algorithm is implemented by using the Stable-Baselines3 library \citep{stable-baselines3} in Python. 

We test the utility of our long-short-term strategy described in Sec.~\ref{rlgeneral} against the traditional long-term strategy. In the latter, the controller is optimized for a long-time horizon of 10 convective time units ($\episodemax=120$) spanning approximately 6 vortex-shedding cycles and weights are updated traditionally, \emph{i.e.} only once in an iteration ($\rltimesteps=120, \multiple=1$ window). On the other hand, in the long-short-term strategy, while the long-time horizon is kept same ($\episodemax=120$), the optimization is performed on two shorter optimization windows ($\rltimesteps=60, \multiple=2$). 

\subsection{Implementation, results, and mechanisms}

\begin{figure}
\centering
\includegraphics[scale=1]{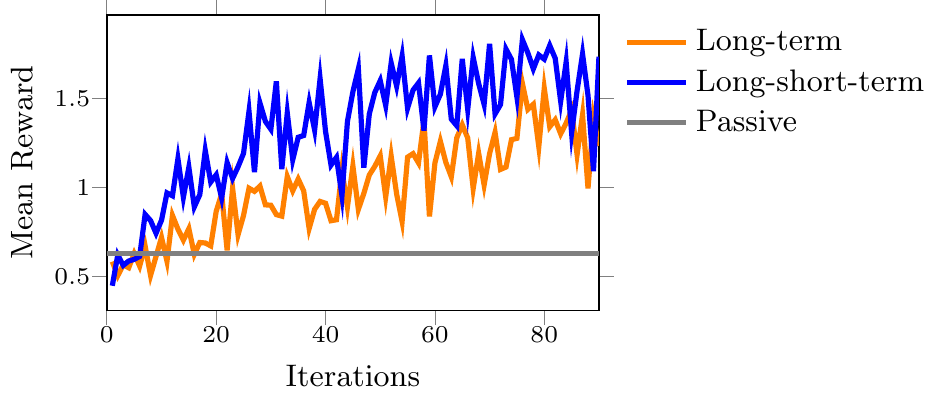}
\caption{Evolution of mean rewards with iterations for different training strategies.}
\label{iterations}
\end{figure}

To demonstrate the effectiveness of the long-short-term strategy, the evolution of the mean reward (sum of rewards divided by episode length) versus iterations for the two learning strategies as well as the passive case of $\stiff=0.015$ (for reference) are shown in Fig.~\ref{iterations}. Firstly, we note that the evolution is oscillatory because of the stochasticity in stiffness sampling during training. 
Next, it can be seen that with increasing iterations, for both cases, the controller gradually learns an effective policy as the mean reward increases beyond the passive reference case. 
However, the long-short-term strategy is found to exhibit faster learning as well as attain a larger reward at the end of 90 iterations as compared to the long-term one.
This is because splitting the long-time horizon into two shorter windows and sequentially updating the weights for each window alleviates the burden of learning an effective policy for the entire long horizon as compared to learning via a single weight update in the long-term strategy.
The remainder of the results focuses on the performance of the control policy obtained after the 90th iteration of the long-short strategy.
A deterministic policy is used for evaluating the true performance of the controller, where the actuation provided by the neural network is directly used as the stiffness instead of stochastically sampling a sub-optimal stiffness in training.

%

\begin{figure}
\centering
\begin{subfigure}[t]{0.95\textwidth}
\centering
\includegraphics[scale=1]{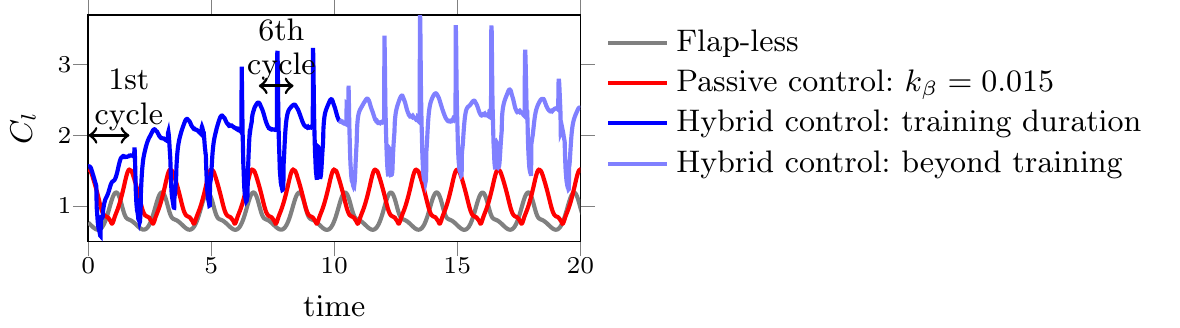}
\vspace{-0.1cm}
\caption{Airfoil lift coefficient.}
\label{liftresult}
\end{subfigure}
\begin{subfigure}[t]{0.48\textwidth}
\centering
\includegraphics[scale=1]{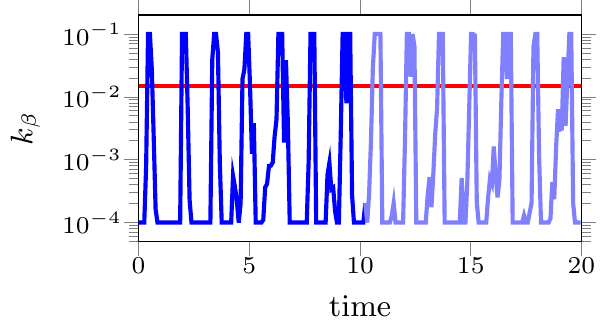}
\vspace{-0.1cm}
\caption{Hinge stiffness.}
\label{stiffresult}
\end{subfigure}
\begin{subfigure}[t]{0.48\textwidth}
\centering
\includegraphics[scale=1]{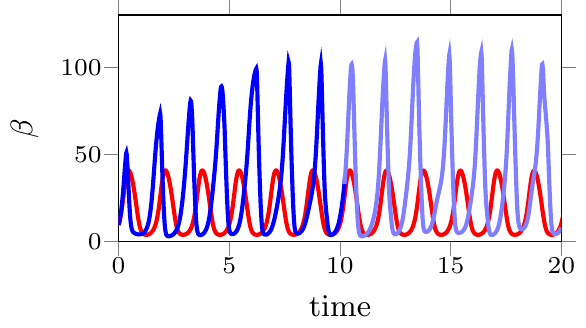}
\vspace{-0.1cm}
\caption{Flap deflection.}
\label{betaresult}
\end{subfigure}
\caption{Temporal plots for flap-less (no control), passive and hybrid flow control cases.} 
\end{figure}

The airfoil lift of the flap-less, maximal passive control and hybrid control cases are plotted in Fig.~\ref{liftresult}. For hybrid control, the lift is plotted not only for $\timet \in[0, 10]$ that the controller has been trained for, but also for $\timet \in[10, 20]$. 
It can be seen that the hybrid controller is able to significantly increase the lift in the training duration and beyond. Overall in $\timet \in[0, 20]$, a significant lift improvement of $136.43 \%$ is achieved as compared to the flap-less case. For comparison, the corresponding lift improvement of the best passive case is $27\%$ \citep{nair2022fluid}. The stiffness actuations outputted by the controller and the resulting flap deflection are plotted in Fig.~\ref{stiffresult} and \ref{betaresult}, respectively. 
It can be observed for hybrid control that the stiffness varies across four orders of magnitude (as compared to fixed $\stiff=0.015$ in passive control) and often reaching its bounding values of $\stiff=10^{-4}$ and $10^{-1}$. Due to these large stiffness variations, the flap oscillates with an amplitude that is more than twice in passive control, indicating that larger amplitude flap oscillations can yield larger lift benefits when timed appropriately with key flow processes. 

\begin{figure}
\centering
\hspace{-0.55cm}
\begin{subfigure}[b]{0.32\textwidth}
\centering
\includegraphics[scale=1]{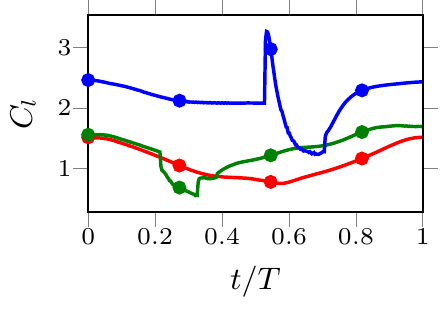}
\vspace{-0.5cm}
\caption{Lift coefficient}
\label{onetcl}
\end{subfigure}
\begin{subfigure}[b]{0.32\textwidth}
\centering
\includegraphics[scale=1]{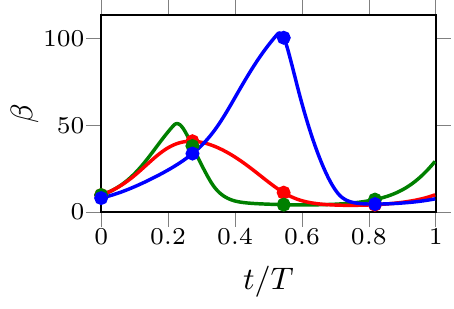}
\vspace{-0.5cm}
\caption{Flap deflection}
\label{onetbeta}
\end{subfigure}
\hspace{0.1cm}
\begin{subfigure}[b]{0.32\textwidth}
\centering
\includegraphics[scale=1]{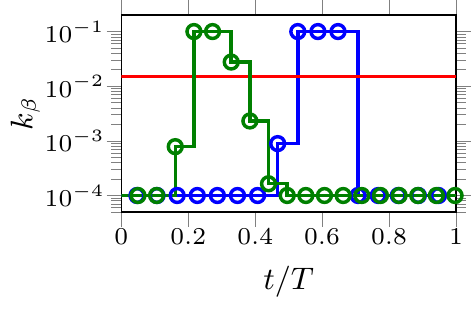}
\vspace{-0.5cm}
\caption{Stiffness}
\label{onetstiff}
\end{subfigure}
\begin{subfigure}[b]{0.35\textwidth}
\centering
\includegraphics[scale=1]{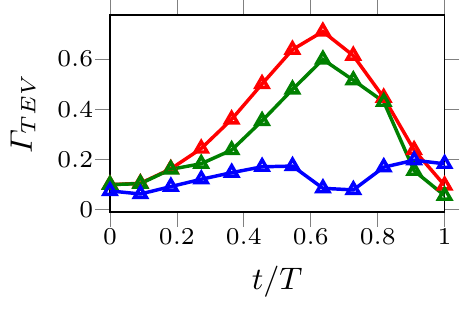}
\vspace{-0.5cm}
\caption{TEV strength}
\label{onettev}
\end{subfigure}
\begin{subfigure}[b]{0.6\textwidth}
\centering
\includegraphics[scale=1]{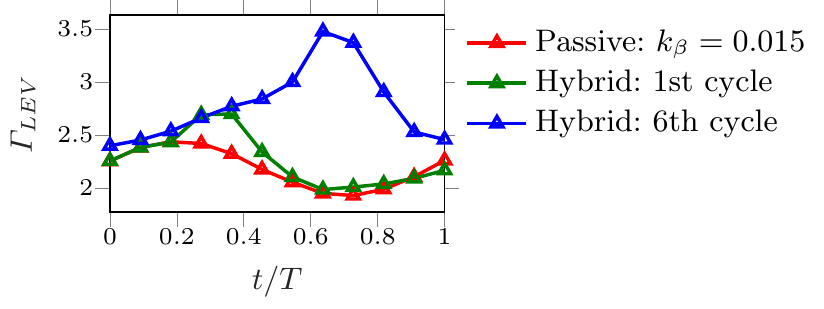}
\vspace{-0.5cm}
\caption{LEV strength (magnitude)}
\label{onetlev}
\end{subfigure}
\caption{Time variation in various quantities during the lone periodic cycle for the passive case, and the first and sixth cycles of the hybrid case (highlighted in Fig.~\ref{liftresult}).}
\label{onet}
\end{figure}

To understand the physical mechanisms driving lift benefits for the hybrid case, we show various quantities during the first and sixth vortex shedding cycles (\emph{c.f.}, Fig.~\ref{liftresult}). These distinct cycles allow for a comparison of the transient and quasi-periodic regimes. The perfectly periodic dynamics of the passive $\stiff=0.015$ case are also shown for reference. 
Firstly, from Fig.~\ref{onetstiff}, we can see that initially in $\periodt\in[0,0.16]$, the controller actuation is lower ($\stiff= 10^{-4}$) than the constant passive actuation ($\stiff=0.015$). This low stiffness prompts the hybrid flap in the first cycle to undergo a slightly larger deflection until $\deflection\approx 50^\circ$ in Fig.~\ref{onetbeta}. 
The decisive actuation occurs at $\periodt\approx 0.21$ when the largest $\stiff=10^{-1}$ is prescribed (\emph{c.f.} Fig.~\ref{onetstiff}), which forces the flap to oscillate downwards within a short time span until $\periodt=0.4$ (\emph{c.f.} Fig.~\ref{onetbeta}). The flap then begins to rise only after the actuation is reduced back to $\stiff=10^{-4}$ by $\periodt=0.5$ (\emph{c.f.} Fig.~\ref{onetstiff}). For comparison, the rising and falling of the single-stiffness flap in the same duration of $\periodt\in[0,0.5]$ occurs gradually (\emph{c.f.} Fig.~\ref{onetbeta}). %
To understand the effect of such an aggressive flapping mechanism on airfoil lift, we plot the circulation strengths of the trailing- and leading-edge vortices (TEV and LEV, respectively) in Fig.~\ref{onettev} and \ref{onetlev}, respectively. Here, $\tev$ and $\lev$ are the magnitudes of positive and negative circulation strengths evaluated in bounding boxes, $[0.85, 1.1]c \times [-0.35, -0.1]c$ and $[0, 1.1]c \times [-0.35, 0.2]c$, respectively. It can be observed that after $\periodt\approx 0.18$ when the flap strongly oscillates downwards in the first cycle, $\tev$ and $\lev$ for the hybrid case are decreased and increased as compared to the passive case, respectively. The overall effect on performance is that the lift of the hybrid case in the first cycle begins to increase at $\periodt\approx 0.4$ after an initial dip as seen in Fig.~\ref{onetcl}. 


\begin{figure}
\begin{adjustwidth}{}{0.5cm} 
\centering
\hspace{-0.45cm}
\centering
\begin{subfigure}[t]{0.281\textwidth}
\centering
\includegraphics[scale=1]{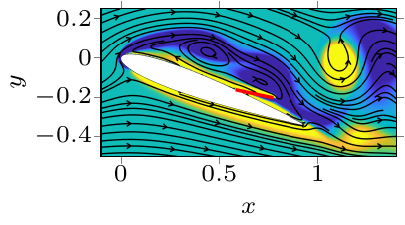}
\vspace{-0.6cm}
\caption{Hybrid: $t=0 \ T$ }
\label{vort61}
\end{subfigure}
\begin{subfigure}[t]{0.22\textwidth}
\centering
\includegraphics[scale=1]{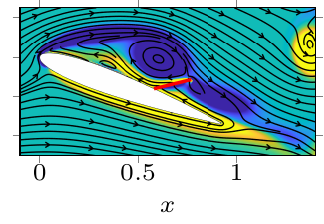}
\vspace{-0.6cm}
\caption{Hybrid: $t=0.27 \ T$ }
\label{vort62}
\end{subfigure}
\begin{subfigure}[t]{0.22\textwidth}
\centering
\includegraphics[scale=1]{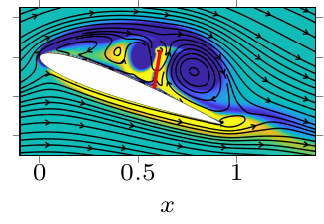}
\vspace{-0.6cm}
\caption{Hybrid: $t=0.55 \ T$ }
\label{vort63}
\end{subfigure}
\begin{subfigure}[t]{0.22\textwidth}
\centering
\includegraphics[scale=1]{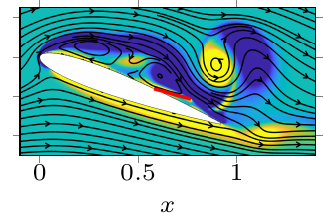}
\vspace{-0.6cm}
\caption{Hybrid: $t=0.82 \ T$ }
\label{vort64}
\end{subfigure}
\centering
\begin{subfigure}[t]{0.281\textwidth}
\centering
\includegraphics[scale=1]{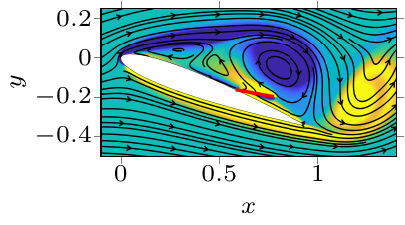}
\vspace{-0.6cm}
\caption{Passive: $t=0 \ T$ }
\label{fvv1}
\end{subfigure}
\begin{subfigure}[t]{0.22\textwidth}
\centering
\includegraphics[scale=1]{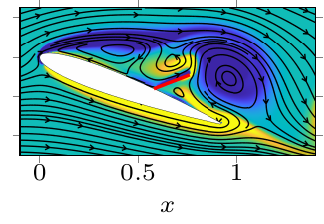}
\vspace{-0.6cm}
\caption{Passive: $t=0.27 \ T$ }
\label{fvv2}
\end{subfigure}
\begin{subfigure}[t]{0.22\textwidth}
\centering
\includegraphics[scale=1]{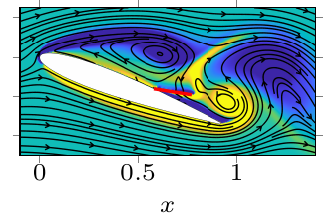}
\vspace{-0.6cm}
\caption{Passive: $t=0.55 \ T$ }
\label{fvv3}
\end{subfigure}
\begin{subfigure}[t]{0.22\textwidth}
\centering
\includegraphics[scale=1]{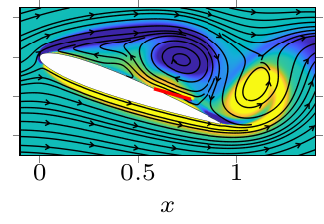}
\vspace{-0.6cm}
\caption{Passive: $t=0.82 \ T$ }
\label{fvv4}
\end{subfigure}
\end{adjustwidth}
\caption{Vorticity contours for hybrid control during the 6th cycle (top row) and passive single-stiffness control (bottom row) at four instances indicated by the markers in Fig.~\ref{onetcl}.}
\label{rlvort}
\end{figure}

Now, as time progresses, this aggressive flapping mechanism continues, but with increasing amplitude as observed in Fig.~\ref{betaresult}. 
Eventually, by the sixth cycle, the switching in stiffness occurs at delayed time instants of $\periodt\approx 0.53$ and $0.7$ (\emph{c.f.} Fig.~\ref{onetstiff}). As a consequence, the flap oscillates upwards until $\periodt=0.5$ and attains a large deflection of $\deflection\approx 100^\circ$ (\emph{c.f.} Fig.~\ref{onetbeta}). Then as the stiffness switches to high $\stiff=10^{-1}$, the flap deflection suddenly drops to $\deflection\approx 5 ^\circ$ within a short time span. Similar to the first cycle, this strong downward motion mitigates the TEV and enhances the LEV as seen in Fig.~\ref{onettev} and \ref{onetlev}, respectively, but now to a much stronger degree. 
To visualize these effects, vorticity contours at four time instants in the sixth cycle are plotted in Fig.~\ref{vort61}--\ref{vort64} and compared to passive control in Fig.~\ref{fvv1}--\ref{fvv4}. The TEV which is clearly decipherable for the passive case in Fig.~\ref{fvv3} is now limited to a much smaller size in the hybrid case in Fig.~\ref{vort63}. This is because the strong angular velocity of the downward oscillating flap sheds away the TEV quickly and restricts its growth. 
This downward motion further contributes to a reduced width of the separated recirculation region in the airfoil-normal direction in Fig.~\ref{vort61} as compared to passive control in Fig.~\ref{fvv1}. Due to the TEV mitigating, LEV enhancing and separation width reducing mechanisms of hybrid flow control, the airfoil attains a much higher lift by the sixth cycle as compared to the passive case (\emph{c.f.} Fig.~\ref{onetcl}).

\begin{figure}
\begin{adjustwidth}{}{0.5cm} 
\centering
\hspace{-0.45cm}
\centering
\begin{subfigure}[t]{0.281\textwidth}
\centering
\includegraphics[scale=1]{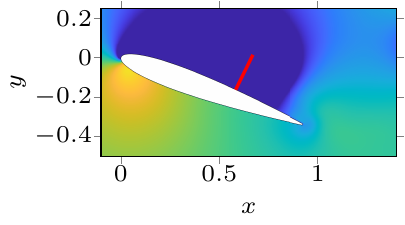}
\vspace{-0.6cm}
\caption{Hybrid: $t=0.45 \ T$ }
\label{cp61}
\end{subfigure}
\begin{subfigure}[t]{0.22\textwidth}
\centering
\includegraphics[scale=1]{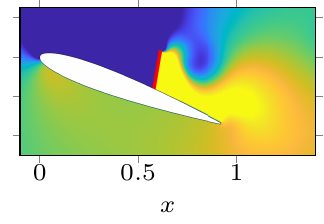}
\vspace{-0.6cm}
\caption{Hybrid: $t=0.55 \ T$ }
\label{cp62}
\end{subfigure}
\begin{subfigure}[t]{0.22\textwidth}
\centering
\includegraphics[scale=1]{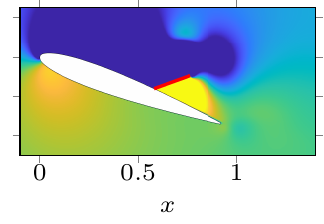}
\vspace{-0.6cm}
\caption{Hybrid: $t=0.64 \ T$ }
\label{cp63}
\end{subfigure}
\begin{subfigure}[t]{0.22\textwidth}
\centering
\includegraphics[scale=1]{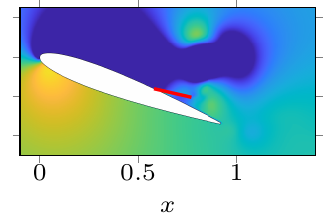}
\vspace{-0.6cm}
\caption{Hybrid: $t=0.73 \ T$ }
\label{cp64}
\end{subfigure}
\end{adjustwidth}
\caption{$\Cp$ contours at four time instants in the sixth cycle of hybrid control.}
\label{rlcp6}
\end{figure}

Finally, we briefly discuss the occurrence of the spikes observed in the lift signal in Fig.~\ref{liftresult} by focusing on the spike in the sixth cycle in Fig.~\ref{onetcl}, visualized via $C_p$ contours at four time instants surrounding the occurrence of the spike in Fig.~\ref{cp61}--\ref{cp64}. 
We note that the spike initiates at $\periodt\approx 0.53$ when the flap attains its highest deflection (\emph{c.f.} Fig.~\ref{onetcl} and \ref{onetbeta}). Preceding this instant, as the flap oscillates upwards, it induces the fluid in the vicinity to move upstream due to its no-slip condition. When the flap comes to a sudden stop as the stiffness switches from $10^{-4}$ to $10^{-1}$ at $\periodt\approx 0.53$ (\emph{c.f.} Fig.~\ref{onetstiff}), the moving fluid in the region post-flap abruptly loses its momentum. This momentum loss is manifested as a strong rise in post-flap pressure (\emph{c.f.}, Fig.~\ref{cp62}). The fluid in the pre-flap region, however, does not experience a barrier in its upstream motion, and instead continues to roll up and builds up the suction pressure. The flap dividing the high-pressure post-flap and low-pressure pre-flap regions can be clearly seen in Fig.~\ref{cp62} and \ref{cp63}, respectively. This large pressure difference across the flap contributes to the large spikes in the airfoil lift. 

\section{Conclusions}

A hybrid active-passive flow control method was introduced as an extension of the covert-inspired passive flow control method consisting of a torsionally mounted flap on an airfoil at post-stall conditions involving vortex shedding. This hybrid strategy consisted of actively actuating the hinge stiffness to passively control the dynamics of the flap. A closed-loop feedback controller trained using deep RL was used to provide effective stiffness actuations to maximize lift. The RL framework was  described, including modifications to the traditional RL methodology that enabled faster training for our hybrid control problem. 
The hybrid controller provided lift improvements as high as $136\%$ and $85\%$ with respect to the flap-less airfoil and the maximal passive control (single-stiffness) cases, respectively. These lift improvements were attributed to large flap oscillations due to stiffness variations occurring over four orders of magnitude. Detailed flow analysis revealed an aggressive flapping mechanism that led to significant TEV mitigation, LEV enhancement and reduction of separation region width. Finally, we remark that since the stiffness changes can be well approximated by a small number of finite jumps, a discrete VSA could be a pathway to realizing this actuation strategy.


\section{Declaration of interests}

The authors report no conflict of interest.

\bibliographystyle{jfm}
\bibliography{references3}

\end{document}